\documentclass[3p,times,twocolumn]{elsarticle}

\usepackage{ecrc}

\volume{00}

\firstpage{1}

\journalname{Nuclear Physics B Proceedings Supplement}

\runauth{}

\jid{nppp}

\jnltitlelogo{Nuclear Physics B Proceedings Supplement}

\usepackage{amssymb}

\usepackage[figuresright]{rotating}

\newcommand{\pp}{$\rm{p}+\rm{p}$}
\newcommand{\pA}{$\rm{p}+\rm{A}$}

\newcommand{\sqrts}{\ensuremath{\sqrt{s}}}
\newcommand{\gev}{GeV/$c$}
\newcommand{\gevtwo}{GeV/$c^{2}$}
\newcommand{\jpsi}{\ensuremath{J/\psi}}
\newcommand{\psitwos}{$\psi(2S)$}
\newcommand{\pT}{\ensuremath{p_{\rm{T}}}}

\begin{document}

\begin{frontmatter}



\dochead{}

\title{Measurement of \jpsi\ production in \pp\ collisions at \sqrts\ = 500 GeV at STAR experiment}


\author{Rongrong Ma, on behalf of the STAR Collaboration}

\address{Brookhaven National Laboratory, Upton, NY 11973, USA}

\begin{abstract}
Quarkonium measurements in heavy-ion collisions play an essential role in understanding the hot, dense medium created in such collisions. As a reference, their production mechanism in \pp\ collisions needs to be thoroughly understood. In this paper, we report the measurement of inclusive cross section of \jpsi\ with transverse momentum (\pT) above 4 \gev\ at mid-rapidity in \pp\ collisions at \sqrts\ = 500 GeV by the STAR experiment. The ratio of the yield of $\psi(2S)$ to \jpsi\ integrated over $4 < \pT < 12$ \gev\ is also presented. Furthermore, the \jpsi\ yields are studied in different event multiplicity bins in different \jpsi\ \pT\ regions, where the low \pT\ measurement is enabled by the newly installed Muon Telescope Detector. A strong increase of the relative \jpsi\ yield with the event multiplicity is observed for all \pT\ with significant \pT\ dependence. 
\end{abstract}


\end{frontmatter}


\section{Introduction}
\label{sect:intro}
In relativistic heavy-ion (HI) collisions, a new state of matter, usually referred to as ``Quark Gluon Plasma (QGP)'', is produced, where quarks and gluons instead of hadrons are the relevant degrees of freedom. Quarkonia have been an important probe to study the properties of QGP as their constituent quarks are predominantly produced in the very early stage of the collision. In particular, \jpsi, a $c\bar{c}$ bound state, is predicted to be suppressed due to the color screening of its binding constituent quarks by the surrounding partons in the medium, which is viewed as a signature of deconfinement \cite{Matsui:1986dk}. Interpretation of the \jpsi\ suppression observed in HI collisions requires thorough understanding of its production mechanism in elementary \pp\ collisions. Despite decades of continuous efforts, understanding of the production mechanism still remains incomplete. Measurements of the \jpsi\ production, including both the inclusive cross section and its dependence on the event multiplicity, at different collision energies could shed light on the understanding of different mechanisms. In particular, the event multiplicity dependence explores the interplay between soft and hard production processes, and could provide new insights into the properties of high-multiplicity \pp\ events.

\section{Detector and data sample}
The Solenoidal Tracker At RHIC (STAR) \cite{Ackermann:2002ad} is a dedicated detector to study HI collisions with coverage over full azimuthal angle and $|\eta|<1$ in pseudo-rapidity. The Time Projection Chamber measures momenta of traversing charged particles with good precision down to low \pT. It also provides ionization energy loss measurement used to identify electrons and muons from \jpsi\ decays. Hadrons can be further distinguished from electrons at high \pT\ with the help of the Barrel ElectroMagnetic Calorimeter (BEMC), in which electrons deposit almost all their energies while the hadrons leave mostly only a minimal amount. On the other hand, muon purity can be further improved using the Muon Telescope Detector (MTD) based on its precise timing measurement as hadrons need more time to reach MTD. Also, the magnet in front of the MTD serves as a hadron absorber. Due to physical constraints, the full MTD covers  about 45\% in azimuth within $|\eta|<0.5$. Only 63\% of the MTD trays were installed in 2013 when the data used in this analysis were taken.

The analysis of reconstructing \jpsi\ via the di-electron channel utilizes the data sample recorded in 2011 for \pp\ collisions at \sqrts\ = 500 GeV. A High Tower (HT) trigger in the BEMC, which requires at least one tower with energy deposition above 3.5 GeV, was used to enhance the statistics for high \pT\ \jpsi. It sampled an integrated luminosity of 22 $\rm{pb}^{-1}$. In 2013, the MTD was commissioned to trigger on muons from heavy flavor decays. Specifically for quarkonium measurements, a dimuon trigger was used which requires at least two MTD hits in coincidence with a MB trigger, and it sampled an integrated luminosity of 28.3 $\rm{pb}^{-1}$ data for \pp\ collisions at 500 GeV.

\section{Results}
\subsection{Signal extraction}
\begin{figure}[htbp]
\centering
\includegraphics[width=0.35\textwidth]{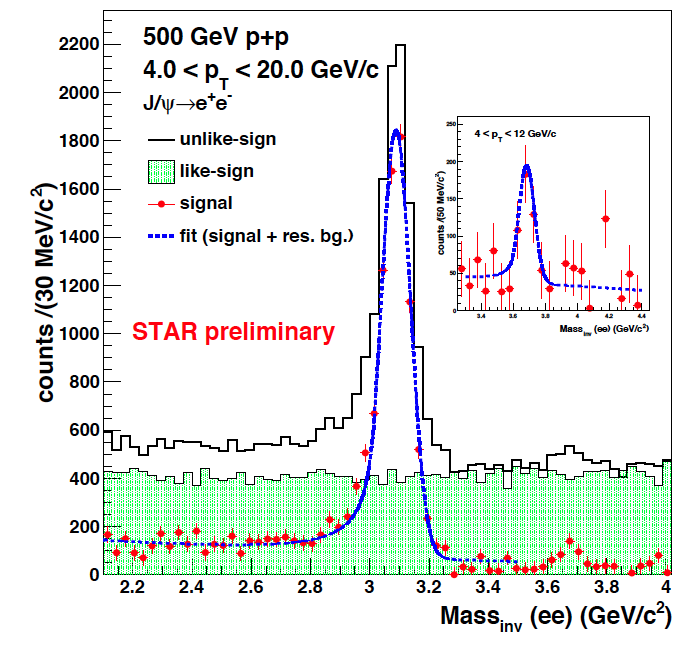}
\caption{Invariant mass distribution of unlike-sign (black histogram) and like-sign (shaded area) pairs of electron and position candidates. The signal distribution (solid circle) is obtained by subtracting like-sign distribution from the unlike-sign. The \psitwos\ signal is shown in the insert as well.
\label{fig:run11-jpsi-signal}
}
\end{figure}
The invariant mass distribution of unlike-sign electron-positron pairs (black histogram) above 4 \gev\ is shown in Fig.\ref{fig:run11-jpsi-signal}, along with a similar distribution for like-sign electron or positron paris (shaded area), which represents a data-driven estimation of the combinatorial background. The signal distribution (red circles) is obtained by subtracting the like-sign distribution from the unlike-sign, and is fitted with a Crystal-ball function describing the \jpsi\ signal shape as well as an exponential function describing the residual background mainly from $c\bar{c}$, $b\bar{b}$ decays and the Drell-Yan process. The raw \jpsi\ yield is obtained by subtracting the integral of the fitted residual background from the sum of the bin contents of the signal distribution in the mass range $2.7 < M_{ee} < 3.3$ \gevtwo. Loss of \jpsi\ counts below the accounted mass regime due to the missing photon in $\jpsi \rightarrow \rm{e}^{+}\rm{e}^{-}\gamma$ decay is estimated to be $\sim 10\%$ using the fitted Crystal-ball function. A similar procedure is used to extract the raw yield of \psitwos\ in the mass range $3.5 < M_{ee} < 3.8$ \gevtwo.

\begin{figure}[htbp]
\centering
\includegraphics[width=0.45\textwidth]{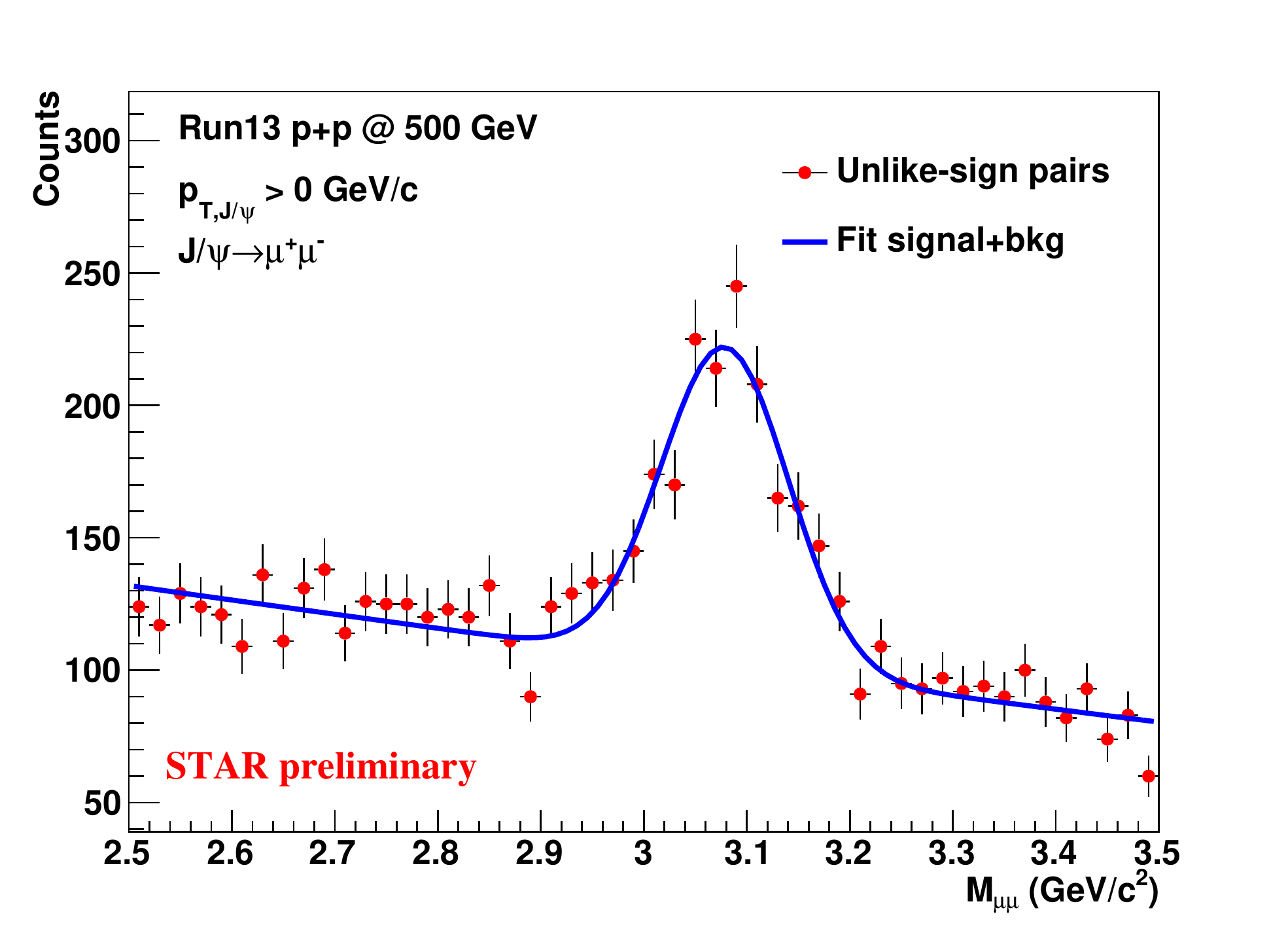}
\caption{Invariant mass distribution of unlike-sign muon pairs (red circles). It is fitted with a Gaussian function describing signal and an exponential function describing background (solid line). 
\label{fig:run13-jpsi-signal}
}
\end{figure}

The invariant mass distribution of unlike-sign muon pairs is shown in Fig. \ref{fig:run13-jpsi-signal} (red circles) using the MTD-triggered data. Contrary to the BEMC trigger, the MTD is able to trigger on low \pT\ muons, which opens access to measurements down to zero \jpsi\ \pT.
The invariant mass distribution is fitted with a Gaussian function describing signal and an exponential function describing background. The raw \jpsi\ yield is extracted by summing the counts of the unlike-sign distribution in the mass range $2.8 < M_{\mu\mu} < 3.3$ \gevtwo\ after removing the background contribution estimated via the fitted function.

\subsection{Inclusive cross section}
\begin{figure}[htbp]
\centering
\includegraphics[width=0.35\textwidth]{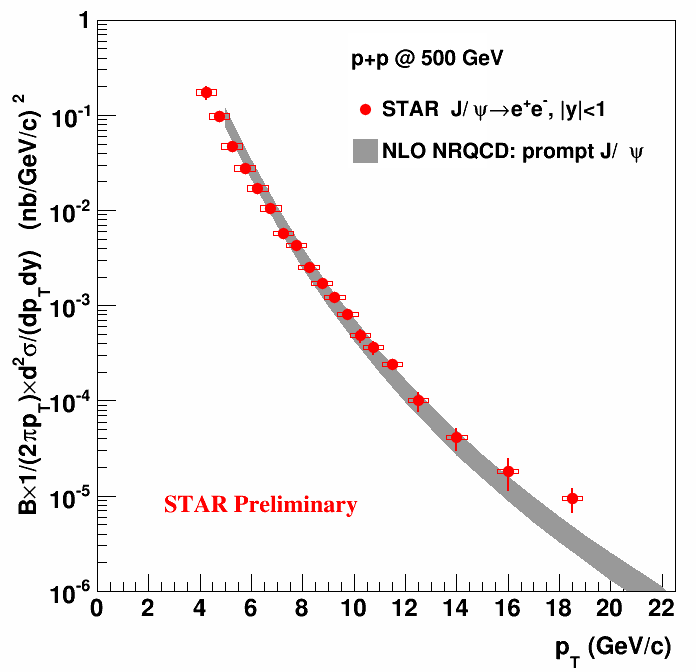}
\caption{Inclusive \jpsi\ cross section as a function of \pT\ at mid-rapidity ($|y|<1$) for \pp\ collisions at 500 GeV measured by STAR (red circles). A NRQCD calculation of prompt \jpsi\ is shown in grey band, and agrees reasonably well with data.
\label{fig:run11-jpsi-xsec}
}
\end{figure}

Figure \ref{fig:run11-jpsi-xsec} shows the inclusive differential cross section for \jpsi\ multiplied by the branching ratio as a function of \pT\ at mid-rapidity ($|y|<1$) measured through the di-electron channel.
The total systematic uncertainty, shown as open boxes around the data points, is about 6.3\%. It includes two main contributions: i) variation of kinematic and electron identification cuts; ii) different mass window and different fitting process to extract signal. A Non-Relativistic Quantum ChromoDynamics (NRQCD) calculation of prompt \jpsi\ \cite{Shao:2014yta} is shown as the grey band in Fig. \ref{fig:run11-jpsi-xsec}. Given that only up to 20\% of  the inclusive \jpsi\ comes from B-hadron decay \cite{Adamczyk:2012ey}, the agreement between the theoretical calculation and the measurement is quite good.

\begin{figure}[htbp]
\centering
\includegraphics[width=0.4\textwidth]{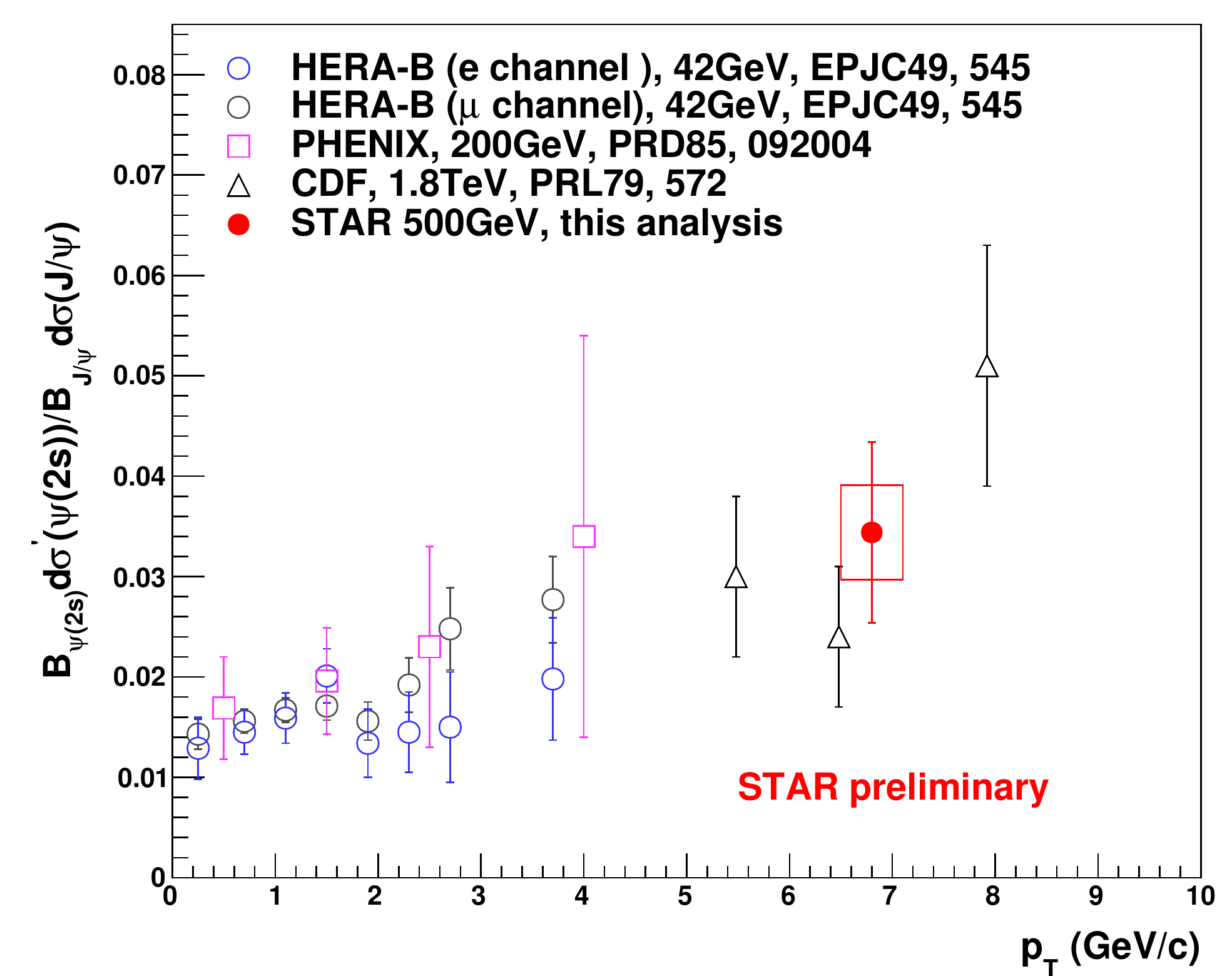}
\caption{Ratio of integrated yield between \psitwos\ and \jpsi\ within $4<\pT<12$ \gev, $|y|<1$. Similar measurements from other experiments are shown for comparison.
\label{fig:run11-jpsi-over-psi}
}
\end{figure}

To help quantify the feed-down contribution of excited charmonium states, the ratio of the integrated yield of \psitwos\ within $4<\pT<12$ \gev\ to that of \jpsi\ is measured and shown in Fig. \ref{fig:run11-jpsi-over-psi}.
It is about 3.5\% at $\pT\approx6.8$ \gev\ with a total systematic uncertainty of about 18\%. Compared with other measurements at different colliding energies for \pp\ or \pA\ systems, the current measurement follows the trend of \pT\ dependence and shows no colliding energy dependence \cite{Abt:2006va,Adare:2011vq,Abe:1997yz}.
 
\subsection{Event activity dependence}
Minimum-bias (MB) \pp\ collisions can be classified into different event activity bins based on the charged-particle multiplicity of the event. To eliminate pile-up effects, charged particles are required to leave a signal in the Barrel Time-Of-Flight (TOF) detector, which is a fast detector insensitive to pile-up. 

\begin{figure}[htbp]
\centering
\includegraphics[width=0.4\textwidth]{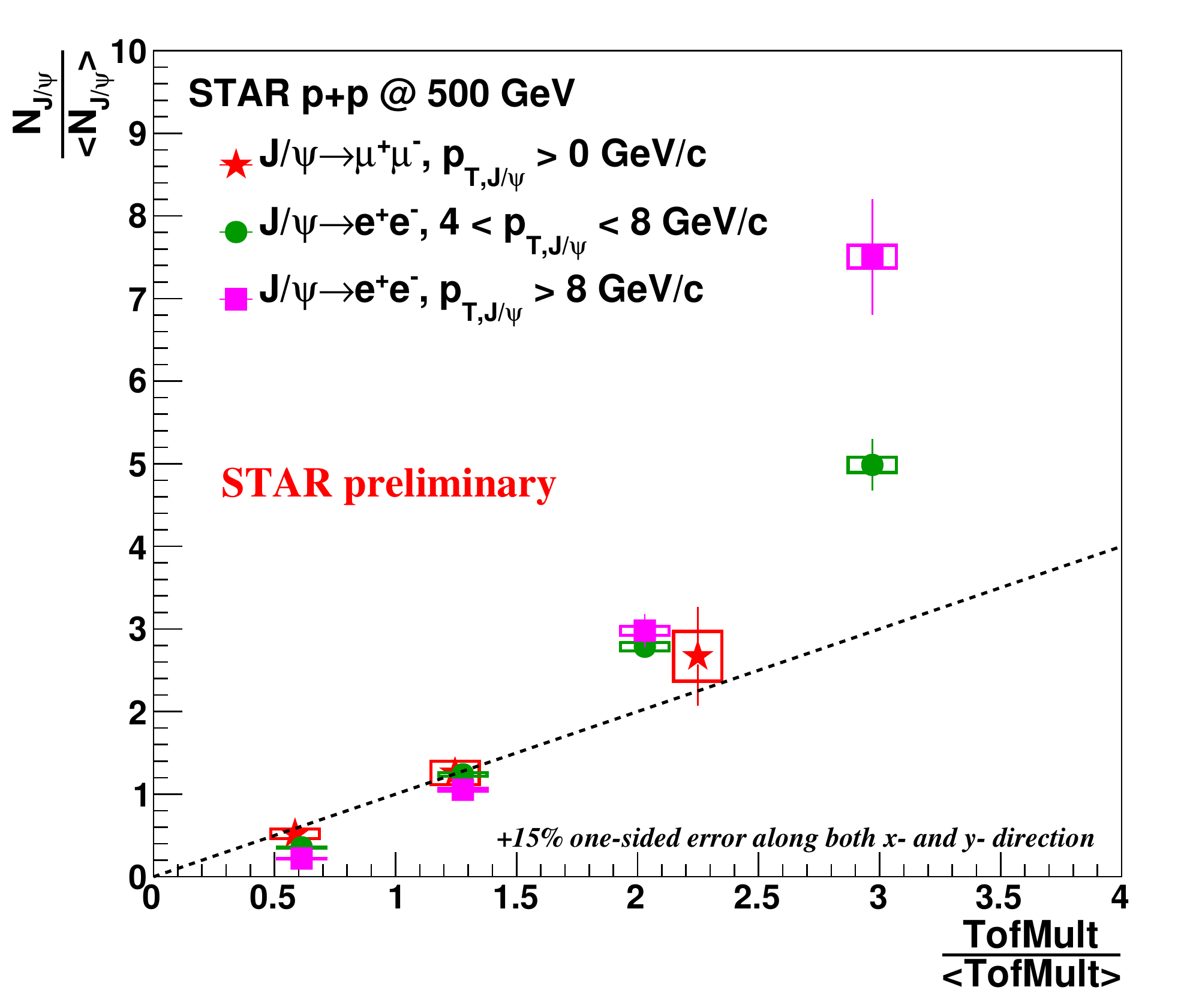}
\caption{Dependence of relative invariant \jpsi\ yield on the event activity in three different \jpsi\ \pT\ bins. The boxes around the data points represent the systematic uncertainties.
\label{fig:evtact-ptbins}
}
\end{figure}

To reduce systematic uncertainties and facilitate comparison to other experiments and to theory, the TOF-matched particle multiplicities and the invariant \jpsi\ yields in different event activity bins are normalized respectively by the same quantity in the MB sample. The resulting dependence of the relative \jpsi\ yield on the event activity is shown in Fig. \ref{fig:evtact-ptbins} in three \jpsi\ \pT\ bins. 
The relative \jpsi\ yield is seen to rise monotonically with the increasing event multiplicity, and the rising trend is steeper for high \pT\ \jpsi. For events with multiplicity about 3 times higher than the average, the relative \jpsi\ yield above 8 \gev\ is enhanced by a factor of about 7.5. For \pT\ dependence, the relative \jpsi\ yield of $4<\pT<8$ \gev\ bin is about 62\% higher than that of $\pT>8$ \gev\ bin in the lowest event activity bin, while it is about 34\% lower in the highest event activity bin. The systematic uncertainty is $\sim11\%$ for $\pT>0$ \gev\ bin from the di-muon channel, and $\sim2\%$ for $4<\pT<8$ \gev\ and for $\pT>8$ \gev\ bins from the di-electron channel. There is an additional one-sided 15\% uncertainty which is due to the lack of reliable simulation to correct for event selection bias in the MB sample.

\begin{figure}[htbp]
\centering
\includegraphics[width=0.4\textwidth]{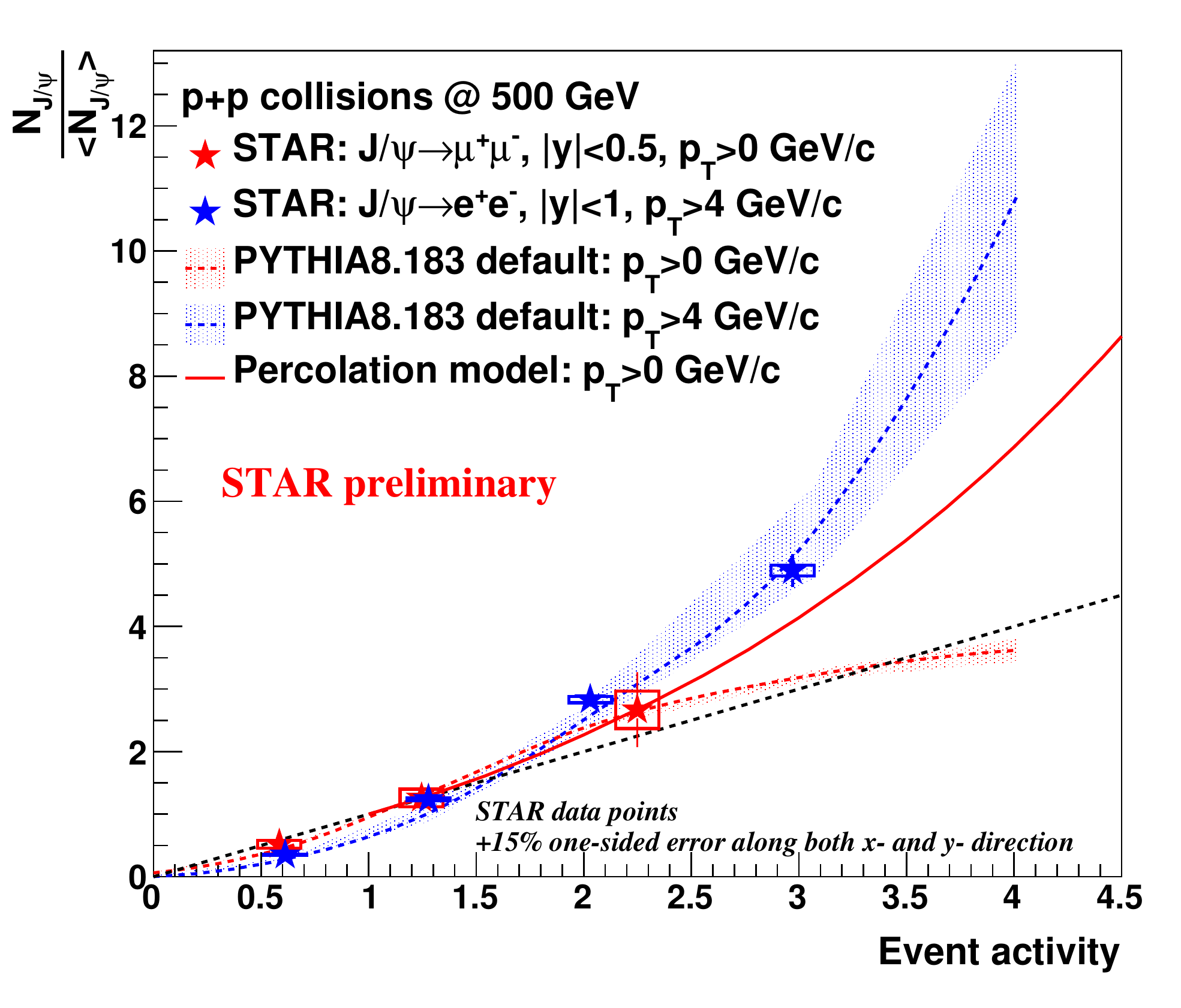}
\caption{Comparison of relative \jpsi\ yield vs. event activity between STAR measurement, PYTHIA8 prediction and the percolation model.
\label{fig:evtact-PYTHIA}
}
\end{figure}

PYTHIA8 \cite{Sjostrand:2014zea} predictions (v8.183, default tune) for \jpsi\ \pT\ above 0 and 4 \gev\ are shown as the dashed lines in Fig. \ref{fig:evtact-PYTHIA}, with the shaded areas representing the statistical errors. They agree quite well with data for both \pT\ bins. The percolation model \cite{Ferreiro:2012fb} also agrees well with data for $\pT>0$ \gev. However, it is worth noting that the percolation model differs significantly from PYTHIA at higher event activity bins, which calls for more measurements.

\begin{figure}[htbp]
\centering
\includegraphics[width=0.4\textwidth]{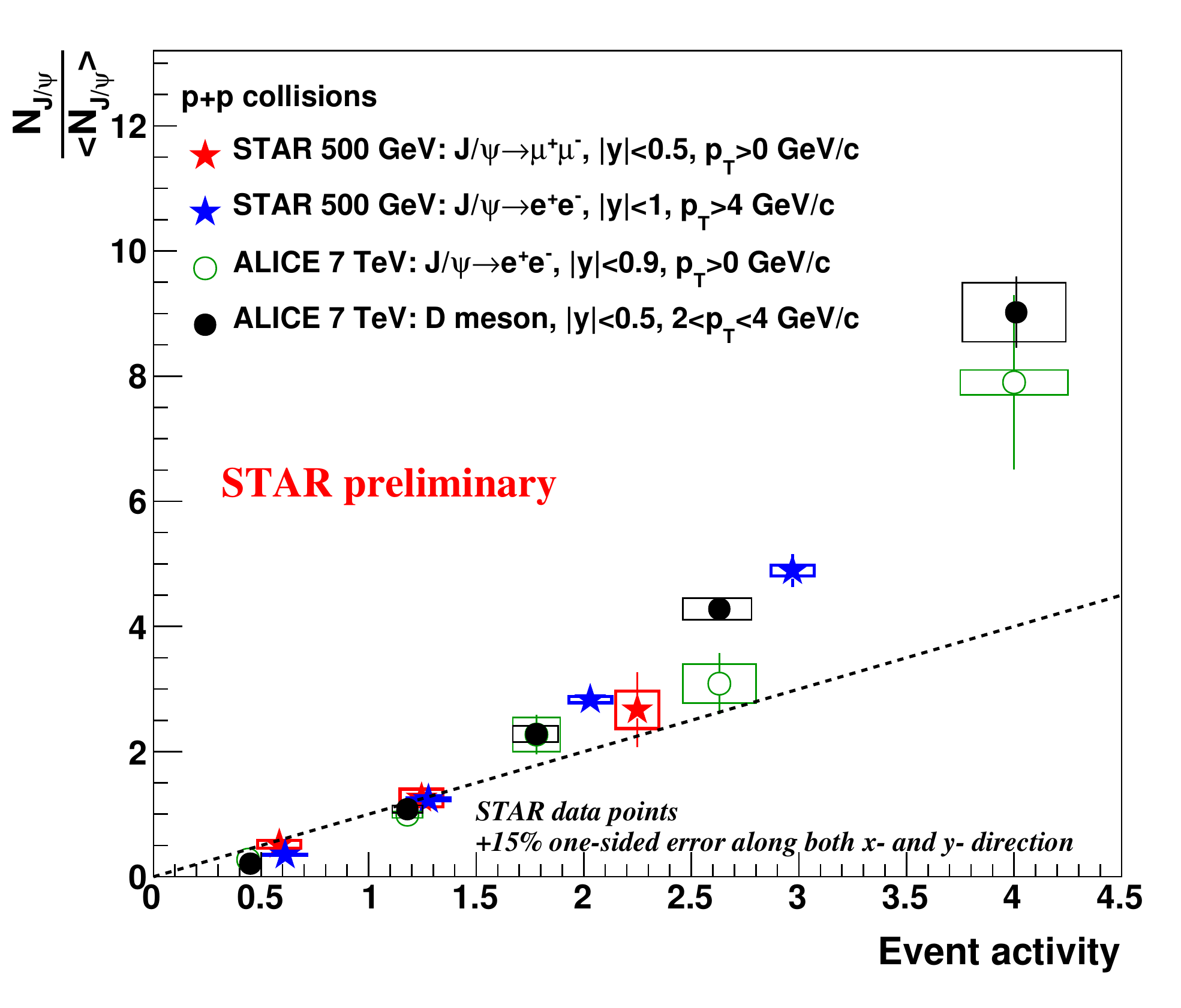}
\caption{Comparison of relative \jpsi\ yield vs. event activity between STAR (500 GeV) and ALICE measurements (7 TeV). A similar measurement of D meson at 7 TeV by ALICE is also shown.
\label{fig:evtact-LHC}
}
\end{figure}

Similar measurements for both \jpsi\ and D meson are observed by the ALICE collaboration for \pp\ collisions at 7 TeV \cite{Abelev:2012rz,Adam:2015ota}, shown as open and closed circles in Fig. \ref{fig:evtact-LHC}.
The stronger-than-linear rising trend at RHIC follows the trend at the LHC quite nicely, which indicates a possible universal dependence of the relative \jpsi\ yield on the event activity at different energies. More differential measurements at different colliding energies are needed to verify this observation. 

\section{Summary}
Differential inclusive \jpsi\ cross-section for transverse momentum above 4 \gev\ is measured at mid-rapidity in \pp\ collisions at \sqrts\ = 500 GeV via the di-electron channel. A NRQCD calculation of prompt \jpsi\ production agrees well with data in the kinematic range where the model is applicable. The ratio of yields of \psitwos\ to \jpsi\ is seen to follow the global trend of \pT\ dependence. Furthermore, the relative \jpsi\ yield is studied as a function of event activity in different \pT\ bins. A strong correlation between \jpsi\ yield and event activity is observed with significant \pT\ dependence. Both PYTHIA and the percolation model can reproduce the data quite well, and the rising trend at RHIC is consistent with that observed at the LHC. More measurements at RHIC are crucial to distinguish different models, and help understand the characteristics of high multiplicity \pp\ events. 



\nocite{*}
\bibliographystyle{elsarticle-num}
\bibliography{Ma_R}







\end{document}